\documentclass[journal,twocolumn,oneside]{IEEEtran}

\usepackage{graphicx}
\usepackage{color}
\usepackage{amsmath}
\usepackage{amsfonts}
\usepackage{epstopdf}
\usepackage{subcaption}

\renewcommand{\eqref}[1]{(\ref{#1})}

\begin{document}

\title{Random Access Protocols for Massive MIMO}
\author{Elisabeth de Carvalho,~\IEEEmembership{Senior Member,~IEEE,}
        Emil Bj{\"o}rnson,~\IEEEmembership{Member,~IEEE,} \\
        Jesper H. S\o{}rensen,~\IEEEmembership{Member,~IEEE,}
        Petar Popovski,~\IEEEmembership{Fellow,~IEEE,}
        Erik G. Larsson,~\IEEEmembership{Fellow,~IEEE}% <-this % stops a space
\thanks{E.~de Carvalho, J.~H.~S\o{}rensen, and P.~Popovski are with the Department of Electronic Systems, Aalborg University, Aalborg, Denmark (email: \{edc, jhs, petarp\}@es.aau.dk). E.~Bj\"ornson and E.~G.~Larsson are with the Department of Electrical Engineering (ISY), Link\"{o}ping University, Link\"{o}ping, Sweden (email: \{emil.bjornson, erik.g.larsson\}@liu.se).This work was performed partly in the framework of the Danish Council for Independent Research (DFF133500273), the Danish National Advanced
Technology Foundation via the VIRTUOSO project, the Horizon 2020 project FANTASTIC-5G (ICT-671660), the EU FP7 project MAMMOET (ICT-619086), ELLIIT, and CENIIT. The authors would like to acknowledge the contributions of the colleagues in FANTASTIC-5G and MAMMOET.}%
}

\maketitle

\begin{abstract}
5G wireless networks are expected to support new services with
stringent requirements on data rates, latency and reliability. One
novel feature is the ability to serve a dense crowd of
devices, calling for radically new ways of accessing the network. This
is the case in machine-type communications, but also in urban
environments and hotspots. In those use cases, the high number of devices and the relatively short channel coherence interval do not allow per-device allocation of orthogonal pilot sequences. This article motivates the
need for random access by the devices to pilot sequences used for channel estimation, and shows
that Massive MIMO is a main enabler to achieve fast access with high
data rates, and delay-tolerant access with different data rate
levels. Three pilot access protocols along with data transmission
protocols are described, fulfilling different requirements of 5G
services.
\end{abstract}

%%%%%%%%%%%%%%%%%%%%%%%%%%%%%%%%%%%%%%%%%%
\section{Introduction}

There is a growing consensus (3GPP, METIS, ITU-R) that 5G wireless
networks will support three generic services:
\begin{itemize}
\item \emph{Enhanced Mobile BroadBand (eMBB)}, with very high data
  rates as the central feature;

\item \emph{Massive Machine Type Communication (mMTC)}, with massive
  numbers of rather simple machine-type devices;

\item\emph{Ultra-Reliable Low Latency Communications (URLLC)}, with
  very low latency and extremely high robustness.

\end{itemize}
Within each category there can be specific services with
additional requirements; e.g. eMBB services that require low
latency are referred to as \emph{Tactile Internet applications}.
While the sheer number of devices is central to the definition of mMTC, it also plays a significant role for eMBB in several challenging scenarios where a large crowd of users is served in a limited spatial region, such as shopping mall, stadium,
or open air festival~\cite{METIS:D11}.  In addition, macro-cells will remain important to provide coverage over larger spatial regions for mobile eMBB services, where dynamic crowds appear along congested streets~\cite{METIS:D11}. 
In this article, we introduce the term {\it cMBB} (crowd
MBB) to denote the distinct class within eMBB related to
crowd scenarios, and describe efficient methods for devices to access
the network in such scenarios. As described below, a key technique is
decentralized assignment of pilot sequences, based on {\it random
  access}. Next, we describe the motivation for random access to
pilots and the distinctive role played by Massive MIMO (multiple-input multiple-output).

\subsection{Why Random Access to Pilots?}

Channel state information (CSI) is necessary for coherent communication.  CSI is particularly important at the
base stations (BSs) in crowd scenarios, since legacy scheduling and
power control algorithms are insufficient to manage many simultaneous
connections. The antenna arrays at the BS are required to manage interference in the spatial
domain through spatial multi-user beamforming, where each beam is
tailored to the CSI of the corresponding device.  The CSI acquisition through pilot sequences is challenging in the use cases with large
numbers of users, cMBB and mMTC.  However, the limitations for CSI
acquisition are different for these two services.  Assuming a simple
protocol with orthogonal pilot sequences, in cMBB the number/duration
of the pilot sequences is limited by the channel coherence time
due to mobility, while in mMTC the devices are generally 
quasi-static and the number of pilot sequences is rather
limited by the uplink power budget.  For mMTC  devices in a 
dynamic environment, mobility may also become a limiting factor. 
Regardless of the reason,
those restrictions put a fundamental limit on the number of pilots
that can be shared by the devices, such that {\it the
  number of devices is much larger than the number of available orthogonal pilot
  sequences}.\footnote{Any number of non-orthogonal pilot sequences can be generated to give each device a unique pilot sequence.
  Instead of being interfered by devices using the same pilot, each device will be partially interfered by a much large set of devices. This interference will be substantial, potentially larger than in the case of orthogonal pilots, thus the use of non-orthogonal pilots does solve the problem. However, it transfers some of the issues from the MAC-layer to PHY-layer
interference mitigation, which is not the subject of this work.}
This is the key motivation random for devising scalable
pilot assignment protocols for cMBB/mMTC.

The wireless traffic specifics are important for choosing the access method for the pilots.  In mMTC, each device is
sporadically active.  In the initial access phase, an active device
connects to the BS, identifies itself and establishes a
coarse synchronization~\cite{Dahlman:2011}.  In principle, the BS can
pre-allocate the pilot sequences to the devices, but this is impractical under  intermittent user activity. The same holds for cMBB in
non-streaming applications, where short periods of activity
alternate with long periods of silence. High device density and
intermittent traffic makes it infeasible to have a dedicated pilot
allocation per device. Instead, we resort to {\it random
access to pilots} (RAP), where a device that wishes to communicate selects randomly a pilot sequence from a predefined
set. RAP may lead to {\it pilot collision}, which is essentially
\emph{pilot contamination}, and the access protocol should deal with it.

\subsection{5G Services and Pilot Shortage}

\begin{figure}[h!]
\begin{center}
\includegraphics[width=\columnwidth]{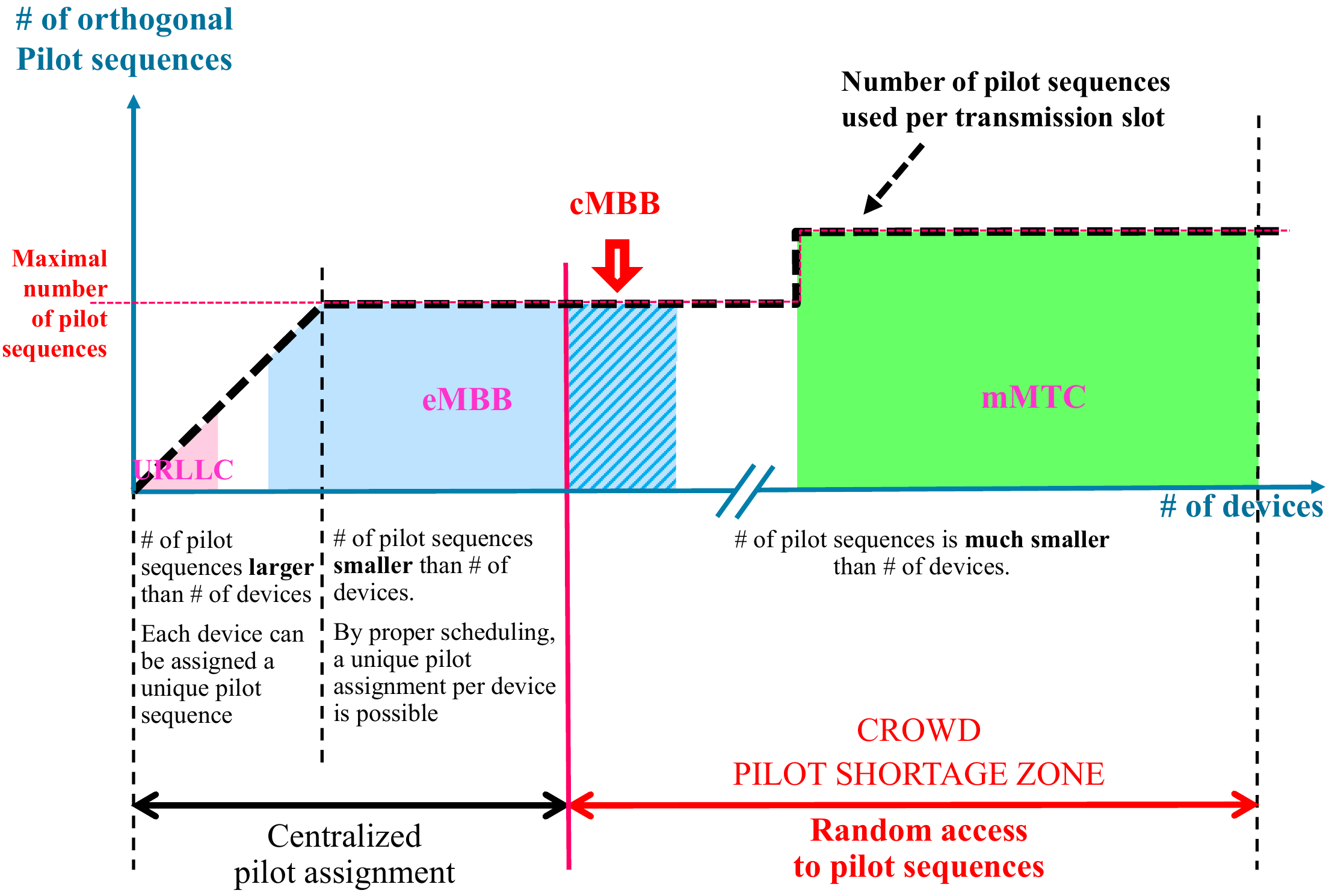}
\end{center}
\caption{ 
Schematic representation of 5G services as a function of the number of devices
  (horizontal axis) and pilot sequences (vertical axis).  When the
  number of devices is not much larger than the 
  number of orthogonal pilot sequences, centralized pilot assignment is possible. In
  a crowd scenario, the number of devices is much larger than the
  number of orthogonal pilot sequences. In this pilot shortage zone
  and with intermittent traffic, random access to pilot sequences
  is a viable solution.
  In the figure, the maximal number of  pilot sequences is arbitrary.
}
\label{fig:5GServices}
\end{figure}

In Figure~\ref{fig:5GServices}, a schematic representation of the 5G
services is provided.  The services are positioned as a function of
the number of devices (horizontal axis) and the number of pilot
sequences available (vertical axis). They are described below.

\begin{itemize}

\item \emph{URLLC:} Due to high reliability requirements, a small
  number of pilots should be exclusively allocated for URLLC. The channel hardening due to large array beamforming mitigates the small-scale fading.
To harness this effect the channel estimation should be precise and
frequent, such that when critical low-latency data arrives, it can be
sent with very high reliability.

\item \emph{eMBB:} The central feature is the high data rate. The number of
  devices is moderate and can be larger than the number of pilot
  sequences. Coordinated pilot assignment remains possible without a
  large overhead. An efficient inter-cell pilot reuse plan can suppress pilot contamination.

\item \emph{cMBB:} The data rate remains as a central feature, but the
  number of devices is much larger than the number of pilots. The
  traffic is intermittent, making the coordinated pilot allocation
  impractical. An option for pilot reuse is to allocate a small set of
  orthogonal pilots to each cell and use large reuse distances.  The
  assignment in each cell is then done by RAP.

\item \emph{mMTC:} The number of devices is in the order of
  10000 or more per BS.  Each device is only sporadically
  active, such that it is necessary to use RAP to connect.

\end{itemize}

The focus of this article is on the random access mechanisms for
scenarios in the pilot shortage zone, as in cMBB and mMTC.
There are three major differences between
cMBB and mMTC in terms of pilot access: (1) the number of devices, which is orders of
magnitude larger in mMTC; (2) the devices in cMBB are located in a small area, while the mMTC are located over a wide area; and (3) the characteristics of the downlink
traffic. In cMBB, traffic volume in the downlink dominates over the uplink, while it is the opposite for mMTC. In this article, we assume that the set of pilots allocated to cMBB is orthogonal to the set of pilots dedicated for mMTC; however, how to multiplex pilots across different traffic types is an open question that warrants further research.

\subsection{Massive MIMO as an Enabler of 5G Services and Random Pilot Access}
\label{sec:MaMIMO}

Massive MIMO, a key ingredient of 5G, can provide very high spectral
efficiency (measured in bit/s/Hz/cell) in sub-6 GHz bands
\cite{Marzetta2010a,MIMO}.  The key idea is to use many antennas at
the BSs, which simultaneously serve many devices through
spatial multiplexing.  Three physical phenomena are important in this
regime.  First, an \emph{array gain} amplifies the signal by focusing it spatially.
Second, \emph{channel hardening} appears, which effectively
eliminates the small-scale fading; that is, each device sees an almost
deterministic (scalar) channel.  Third, the
\emph{high spatial resolution} improves the ability to separate users
spatially and facilitate spatial multiplexing to many users
simultaneously.  Effectively, each device gets an exclusive, focused
data beam and does not suffer from neither small-scale fading nor
interference. The interference between users that utilize the same
pilot is, however, hard to suppress, resulting in pilot contamination \cite{Marzetta2010a}.

Random access in Massive MIMO benefits from the three aforementioned phenomena \cite{Bjornson2016e,deCarvalho2016a}. The array gain
improves the signal-to-noise ratio (SNR), and hence detect weak devices. Channel
hardening facilitates the application of algorithms that exploit
asymptotic channel properties. The multiplexing capability
offers the possibility of \emph{spatially} resolving collisions, an
entirely new opportunity as compared to legacy systems \cite{Hasan2013a}.

The introduction of Massive MIMO on the physical layer requires a
paradigm shift at the MAC layer.  Several essential
assumptions that underpin resource allocation algorithms in 3G or 4G
become questionable. For example, user scheduling becomes unnecessary
due to the fact that every user in Massive MIMO can utilize the full
bandwidth and through a nearly deterministic 
channel, while being spatially separated from the other users. This calls for new MAC design for eMBB/cMBB in order to unleash the full potential of Massive MIMO. For mMTC, a MAC-layer redesign is
required since these services pose entirely new problems, which
cannot be handled with legacy networks or small-scale MIMO
technology.

In the sequel, we  give an overview of the mechanisms to access a
limited set of pilots in cMBB and mMTC, including new approaches.  
Pilot selection is performed at the device, and CSI is acquired at the BS based
on the uplink pilots.  A time-division duplexing (TDD) system is
targeted where channel reciprocity is used to obtain downlink CSI
estimates directly from the uplink CSI estimates.

%%%%%%%%%%%%%%%%%%%%%%%%%%%%%%%%%%%%%%%%%%
\section{Two Classes of Transmission Methods based on Random Access}
\label{sec:classes}

The access methods are described using a terminology that is derived from  conventional random access. 
When a device wants to access the wireless network, it randomly selects an \emph{access sequence}, which is a pilot sequence that can be used for channel estimation, but also to request a pilot sequence for a subsequent collision-free transmission. 
We say that a \emph{collision} occurs if more than one device transmit  the same pilot sequence simultaneously. 
If multiple devices are involved in a collision, but the channel of one or more of them can be estimated and their identity can be determined,
then a \emph{capture} occurs. 
Specifically, if the capture of a device occurs due to its favorable spatial position with respect to the other devices relative to the BS array, then we refer to it as a \emph{spatial capture}. 
If the BS or device is capable of detecting a collision, then it can start a \emph{collision resolution} process by explicitly sending messages that govern the future (randomized) action of the accessing UEs. 

The schemes described here are based on slotted transmission, where a slot is limited by the channel coherence time and bandwidth. 
We assume that the initial timing mismatches in the cell are within the cyclic prefix (CP), such that the orthogonal access sequences remain orthogonal when received at the BS.\footnote{This assumption is valid for a cell radius of 750 m with an LTE-type normal CP, while a cell radius of 2.5 km is supported by the extended CP.} Two classes of approaches are distinguished: 

\begin{itemize}

\item {\bf Random access to pilots (RAP)}. 
Here random access is performed for the sole purpose of being granted a  pilot sequence that can be used in a collision-free transmission. 
A special set of non-dedicated access sequences are used and collisions happen in the pilot domain. 
When a pilot access is collision-free, the corresponding device can be identified, admitted to the network, and assigned a cell-unique pilot. It is henceforth allowed to transmit and receive data, without intra-cell pilot contamination. 
Collision-free access is enabled by a mechanism that is iterative in general,  implying multiple transmission phases between the BS and the devices. RAP finds its primary application in two cases. 

\begin{itemize}
\item The data size is sufficiently large to justify the overhead of random access. 
\item  The traffic is delay-tolerant (a norm in mMTC), but the data volume per mMTC device should be larger than a threshold (not determined here) to justify the access overhead.
\end{itemize}

\item {\bf Random access to pilots and data transmission: RAPiD}.
These are based on random access to the pilot sequences, followed by uplink data transmission. Collisions happen in the pilot domain, while collision-induced interference happens in the data domain.
The RAPiD transmission schemes rely on multiple-slot transmission with pilot hopping across the slots. Pilot hopping brings diversity, since that the data from a given device will be affected by different contamination events across the transmission slots. 
A mechanism is needed to identify the transmitting devices (see Sections~\ref{sec:ERAPiD} and~\ref{sec:CRAPiD}).
Two transmission schemes are later described. The scheme in Section~\ref{sec:ERAPiD} is for delay-tolerant applications with uplink dominant traffic, while the scheme in Section~\ref{sec:CRAPiD} is for delay-stringent applications.

\end{itemize}

\begin{figure*}[t]
        \centering
        \begin{subfigure}[b]{1.8\columnwidth} \centering 
                \includegraphics[width=\columnwidth]{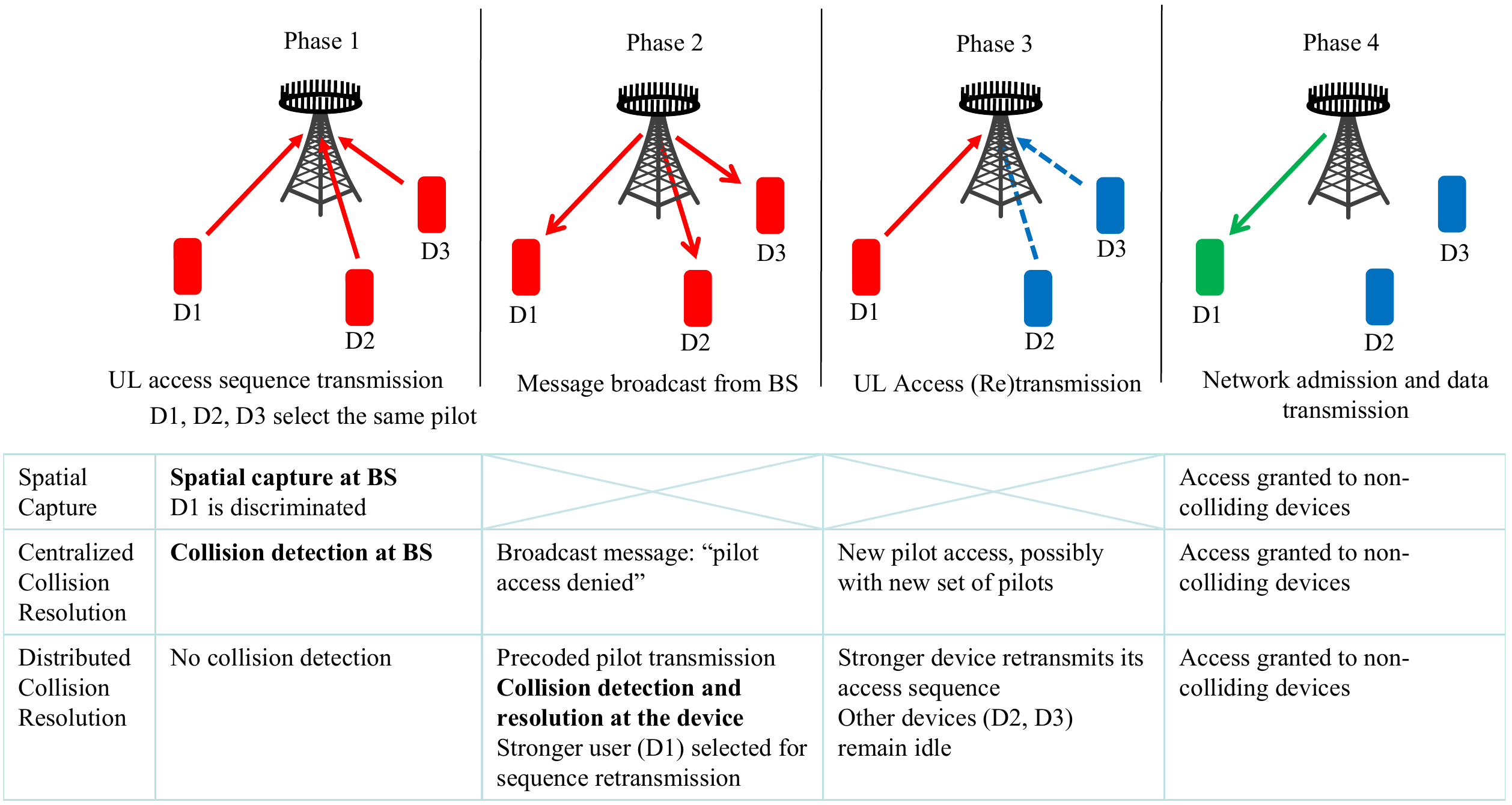} 
                \caption{}
                \label{figure:Classa}
        \end{subfigure} \\ [10mm]
        \begin{subfigure}[b]{1.3\columnwidth} \centering
                \includegraphics[width=\columnwidth]{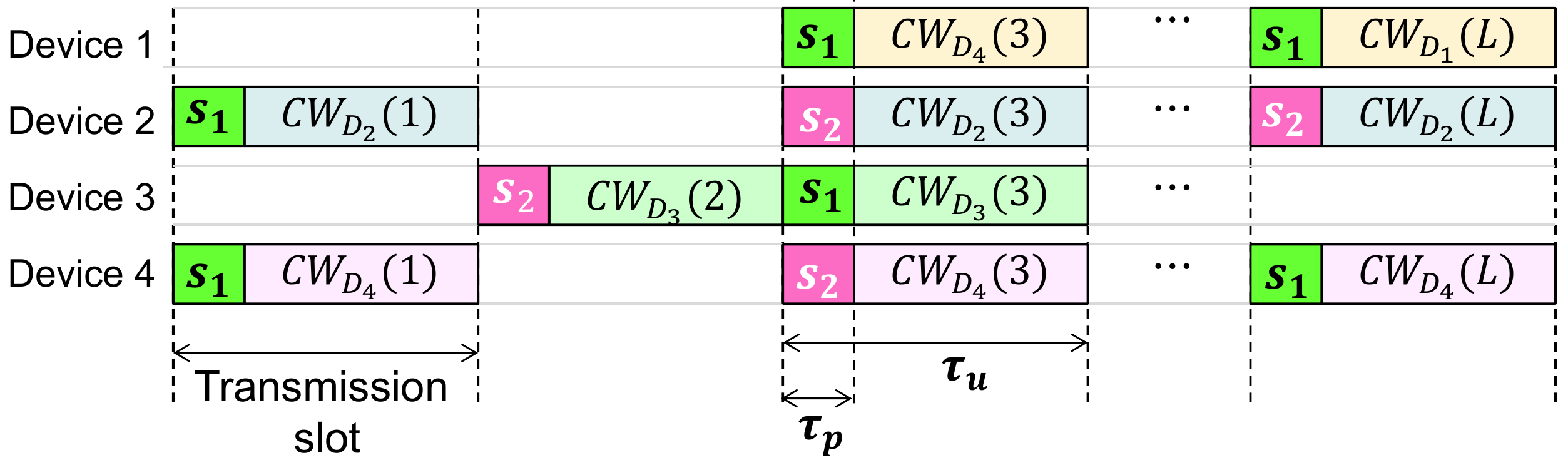} 
                \caption{}
                \label{figure:Classb} 
        \end{subfigure} 
        \caption{Two classes of random access protocols. 
        (a) Random access to pilots (RAP) 
        (b) Random access to pilots and data transmission (RAPiD).: simplified example with four devices and two pilot sequences $\{ {\mathbf s}_1, {\mathbf
            s}_2\}$.  $CW_{D_i}(j)$ denotes the portion of the codeword transmitted by device $i$ during transmission slot $j$. }
\end{figure*}

\section{Random Access to Pilots (RAP)}
\label{sec:RAP}

Figure~\ref{figure:Classa} provides a generic depiction of the RAP protocol for  three methods described in this section. 
The first method takes two transmission phases, while the other two take four transmission phases. 
The transmission phases are as follows:
\begin{itemize}
\item Phase 1: Uplink access sequence transmission.
Each active device selects an access sequence uniformly randomly from a predefined set. 
In addition, a message allowing collision detection at the BS might be transmitted. 

\item Phase 2: Message broadcast from BS to colliding devices. 
For the centralized collision resolution method, this message indicates that a collision was detected.
For the distributed collision resolution method, a precoded pilot signal is transmitted enabling distributed collision resolution within the group of colliding devices.

\item  Phase 3: Uplink pilot retransmission. 
For the centralized resolution method, a new random access sequence is transmitted, possibly from a different set of pilot sequences. 
For the centralized resolution method, a distributively selected device retransmits the access sequence used in Phase 1. 

\item  Phase 4: Network admission message to non-colliding devices and possibly downlink data.
\end{itemize}

\subsection{Spatial Capture}
\label{sec:methodA}

When there is a collision in Phase 1, the BS can use the received signal to separate the devices, utilizing the high spatial resolution of massive MIMO.
In \cite{Sanguinetti2016a}, spatial capture is enabled by timing mismatches between the devices in an OFDM system. The access channel consists of $N$ consecutive subcarriers over which orthogonal access sequences are transmitted in parallel. 
In the frequency domain, a timing offset gives each user a unique signature that spreads over the $N$ subcarriers of the access channel, similarly to the spatial signature of a transmitting source impinging on a uniform linear antenna array of size $N$.
The sample covariance matrix is computed by averaging over the antennas and used to determine the number of colliding devices and the corresponding timing offsets. 
These are then used to estimate the respective channels, which the BS uses in Phase 4 to grant access. The estimation only works if the colliding devices have resolvable timing offsets and 
their number less than $N$. 

Subspace-based methods based on the covariance matrix of the received signal at the BS can also be employed for spatial capture.  In a model with pilot transmission, the channels of the colliding users cannot be distinguished based on the pilot signals.  Hence, subspace-based methods assume data mode transmission where the transmitted data from each device  is sufficiently different to allow channel separability. As blind methods leave an ambiguity in the multi-device channel estimation, additional training is required. These  methods are  sensitive to the estimation quality of the sample covariance matrix as well as the number of devices. In  \cite{Mueller2014b}, the difference of received power between devices in a cell and contaminating devices in neighboring cells is exploited to separate the associated subspaces. This distinction cannot be done in our framework, as the colliding devices belong to the same cell and have comparable received power levels.

\subsection{Centralized Collision Resolution}
\label{sec:methodB}

Centralized collision resolution assumes the existence of a mechanism  at the BS  to detect whether a collision occurred or not. 
The collision detection mechanism in~\cite{Caire2016}  can be adapted to our framework. 
Pilot access in~\cite{Caire2016} does not rely on random access but rather on an aggressive pilot reuse plan that causes pilot collisions in a distributed network. 
Collision detection at the BS is enabled by coded pilot sequences. 
The coded pilot sequence has two parts: (1) non-zero symbols used for channel estimation (useful part) and (2) $l$ null symbols used for collision detection and are placed at random positions in the sequence. 
Each device is identified via a unique on-off pattern.
If, at reception, the number of silent symbols is smaller than $l$, it means that  a collision occurred.  
A new pilot access is then performed in Phase 3 and the probability of a new collision event is reduced.
If the number of  silent symbols is equal to $l$, then no collision occurred. 
At last, note that subspace-based methods from Section~\ref{sec:methodA} also allow for collision detection.

\subsection{Distributed Collision Resolution}
\label{sec:methodC}

Here the access collisions can be detected at each device. The BS uses the received signal from Phase 1 for estimating the sum of the channels that the access sequence propagated over. This estimate is used to send a precoded response in Phase 2, which becomes multicasted towards the colliding devices. In case of no collision, one device receives a precoded signal in Phase 2 that is $M$ times stronger compared to the reception when the BS sends in broadcast mode. The device can measure the array gain reliably due to the channel hardening. If, instead, two devices collide, with path-gains $\beta_1$ and $\beta_2$, 
they can measure the individual array gains $\frac{\beta_1}{\beta_1+\beta_2} M$ and $\frac{\beta_2}{\beta_1+\beta_2} M$, respectively, which sum up to $M$. The first device can distributively detect the collision by checking if $\frac{\beta_1}{\beta_1+\beta_2} M < M$ and similar for the second device. This procedure can be applied with any number of colliding devices.
  
The estimated individual array gains can be used for distributed collision resolution, by setting a criterion on when a device may repeat its pilot in Phase 3. For example, the ratio $\frac{\beta_1}{\beta_1+\beta_2}$ informs Device~1 how strong its path-gain is compared to the contenders. The strongest-user collision resolution (SUCRe) decision criterion was proposed in \cite{Bjornson2017a}, where only the device with the strongest path-gain repeats the pilot in Phase 3. In a two-user collision, Device 1 can be sure to be strongest if $\frac{\beta_1}{\beta_1+\beta_2}>0.5$. The SUCRe protocol exploits the natural variations in path-gains that occur due to different propagation distances and shadowing, in contrast to LTE that attempts to mitigate these variations by power control.

Numerical results have shown that the SUCRe protocol can distributively resolve 90\% of the collisions. When there is a collision-free pilot transmission in Phase 3, the BS grants access to the corresponding device in Phase 4. The remaining collisions can be handled by repeating the protocol after a random waiting time and/or applying collision resolution and spatial capture at the BS.

\begin{figure}[t]
 \centering
 \includegraphics[width=1\columnwidth]{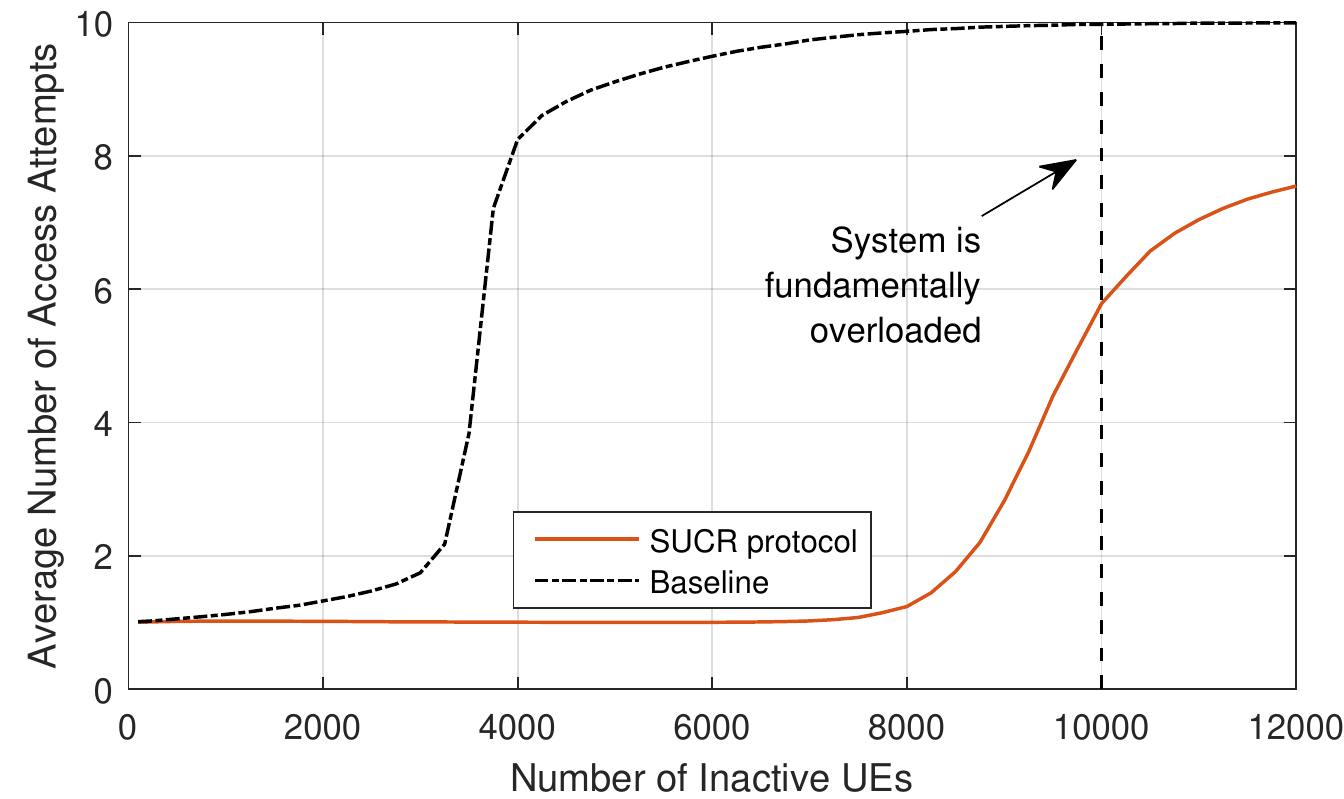}
 \caption{In the method "distributed collision resolution", access collisions can be  resolved in a distributed manner, even under very high load.}
 \label{figure:SUCRprotocol}
\end{figure}

The benefits of this method are illustrated in Figure~\ref{figure:SUCRprotocol} for a crowd scenario with $K \in [100,12000]$ devices, $M=100$ antennas, and $\tau_p = 10$
access sequences. The devices are uniformly distributed in a hexagonal cell
and each one decides to access the network with $0.1\%$ probability, which corresponds to one access per minute if the protocol is repeated every 60 ms. The average cell-edge SNR is 0 dB. If a device is not admitted immediately, it makes a new attempt at the next access occasion with probability $0.5$. After 10 failed attempts, the access is
considered denied. Both numbers can be optimized for a given scenario.

Figure~\ref{figure:SUCRprotocol} shows the average number of
attempts per device, as a function of $K$.  The SUCRe protocol handles
up to $K=8000$ devices without noticeable delays. Notice that
with $K=10000$ there is on average $K \cdot 0.001 / \tau_p =1$ device
per access sequence, meaning that the network is fundamentally
overloaded. Nevertheless, the average access delays are small because an
astonishing $90\%$ of the devices are still admitted to the
system. This behavior remains also for $K>10000$. Figure~\ref{figure:SUCRprotocol} also shows a baseline protocol
where pilot collisions are not resolved at the devices, but by
making new attempts after random time instants. This protocol
requires many more access attempts and breaks
down at around $K=3000$. With $K=10000$ only $1.5 \%$ of the
devices are ever admitted.
Note that we do not compare the SUCRe protocol with the spatial capture scheme from \cite{Mueller2014b} since these aims at two fundamentally different scenarios: small timing mismatches and large timing mismatches, respectively.

\section{Random Access to Pilots and Data Transmission (RAPiD)}

Two schemes pertaining to the RAPiD protocol are described on Figure~\ref{figure:Classb} and both rely on multiple-slot transmission and pilot hopping. In the first scheme, codewords are transmitted  over  multiple transmission  slots so that contamination is averaged out.  
In the second scheme, a given device retransmits the same data content in each transmission slot, according to the activation probabilistic model; this is done until  the pilot transmission is collision-free. 

\subsection{E-RAPiD: Averaging of the Contamination}
\label{sec:ERAPiD}

The basic idea of this protocol is to randomize the effect of
pilot contamination over multiple transmission slots.              
Figure~\ref{figure:Classb} depicts a simplified example, where  $\tau_u$ is the 
duration of a transmission slot. We assume a block fading model, with 
 independent realization in each slot and for each device. The
protocol relies on three main features:
\begin{itemize}
\item {\it Pilot hopping}: In each transmission slot, each active device
  randomly selects one sequence from the
  set of orthogonal pilot sequences. Note that the CSI needs
  to be estimated at the BS in each transmission slot.

\item {\it Ergodic data transmission}: For each device, the codeword
  is divided into multiple parts,
  sent after the pilot sequence. 

\item {\it User discrimination}: From a single transmission slot it is
  impossible to discriminate between colliding devices.  However, the
  whole series of pilots selected by a device across all transmission
  slots provides a unique identifier that is used for decoding.

\end{itemize}
For an asymptotically large number of transmission slots, one codeword is
affected by an asymptotically large number of channel realizations and
interference events.  Interference includes the interference caused by
pilot contamination: for a given device, in each of its active transmission
slots, contamination comes from a different random set of devices, so that
asymptotically the device is affected by all possible sets.  
Depending on the type of receiver, interference 
might also come from
devices that use different pilots. The ergodic
properties of this transmission process achieve 
 a reliable data rate. The transmission protocol is
called \textit{Ergodic Random Access to Pilot and Data transmission}
(E-RAPiD).

\begin{figure} 
 \centering
        \includegraphics[width=\columnwidth]{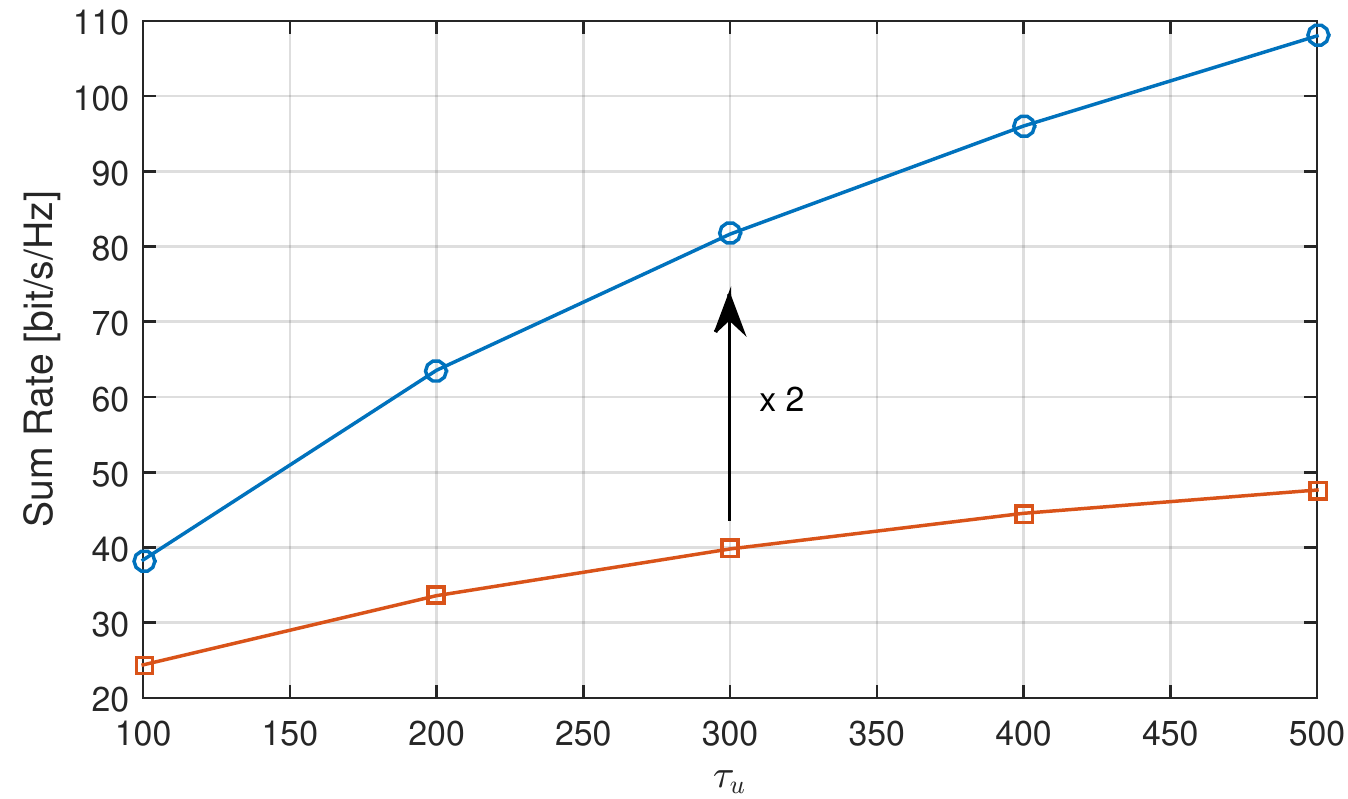} 
        \caption{The lower bound ${\cal R}$ on the sum rate for $K=800$, for $M=100$ and $M=400$
          antennas as a function of the transmission slot duration.}
\label{figure:ERAP}
\end{figure}

The performance of E-RAPiD is characterized using a lower bound on the
uplink sum rate that accounts for the probabilistic user activity~\cite{deCarvalho2016a}.
The bound ${\cal R}$ depends the total number of BS antennas $M$
and the number of pilot sequences, $\tau_p$: the larger those quantities, the more devices can be multiplexed. 
${\cal R}$ is also a function of the device activation probability, $p_a$.  
To maximize the sum rate, one can optimize $p_a$ and $\tau_p$. 
Heuristics leads to the following approximate solution:
\begin{itemize}
\item One third of the transmission slot is used for pilots.

\item The average number of active users is equal to $x \sqrt{M
  \tau_u}$, where $\tau_u$ is the slot duration and  
   the scalar $x$ depends on the distribution of device channel path-gain.
  The larger the variance, the larger $x$ and hence active users. 
\end{itemize}
This solution leads to a scaling of the sum rate as $\sqrt{M \tau_u}$.

Figure~\ref{figure:ERAP} shows  bound ${\cal R}$, optimized w.r.t   $p_a$  and $\tau_p$ for a scenario with 
$K=800$ devices and as a function of the transmission slot duration $\tau_u$. 
Maximum ratio combining is applied to the BS. 
The channel
path-gains vary uniformly at random around a fixed value $\bar{\beta}$
with a maximal gap of $0.25 \bar{\beta}$. The performance is limited by
interference and hence is quite insensitive to the SNR.
{For $\tau_u = 300$, 
the average number of active devices is around 60 for $M=100$ and around 140 for
$M=400$, so that the average rate per active device is equal to 0.5 bit/s/Hz for
both $M=100$ and $M=400$.}

\subsection{C-RAPiD: Intra-cell Interference Cancellation}
\label{sec:CRAPiD}

The third access protocol targets mMTC scenarios, like the E-RAPiD protocol. It also uses pilot hopping for joint pilot and data transmission. However, instead of spreading the codewords across multiple transmission slots, the codewords are replicated in each transmission slot within a predefined duration, called a frame. The same technique is applied in works on coded random access \cite{PSLP2014,Liva2011}. The duration of a frame, $\Delta$, is in general lower than ergodic transmissions require, hence this protocol targets more delay-sensitive applications in the mMTC category. The main idea is to use the successful decoding of a data transmission in one slot to cancel the interference brought by replicas in other slots. While sending multiple replicas increases the intra-cell pilot contamination, it also provides multiple opportunities to successfully decode the codeword. This is the key trade-off in the \textit{Coded Random Access to Pilot and Data transmission} (C-RAPiD) protocol, such that the careful selection of $p_a$ is very important.

The C-RAPiD protocol consists of three steps~\cite{Sorensen2016a}:

\begin{enumerate}
\item \textbf{RAPiD access:}
All $K$ users participate in a RAPiD access procedure, as described in Section~\ref{sec:classes}, for $\Delta$ transmission slots. In each slot, each user is active with probability $p_a$. An active user selects randomly one of the $\tau_p$ pilots and transmits the pilot and a replica of its data message.

\item \textbf{Maximum ratio combining (MRC):} The received signals are processed using MRC based on the contaminated estimates achieved from the pilot transmissions. MRC transforms the signals from linear combinations with small-scale fading coefficients to linear combinations with large-scale fading coefficients (the Euclidean norm of the channel coefficients). Note, that this transformation is only possible due to the channel hardening and beam decorrelation brought by Massive MIMO.

\item \textbf{Successive interference cancellation (SIC):} The linear combinations achieved through MRC represent a system of equations, solved through SIC. Initially, the BS locates immediately decodable data, which is practically done through error checking codes. Embedded in the data is the random activity and pilot choices of the device, which allows the BS to locate all the replicas of the same packet. This enables the BS to cancel the interference caused by these replicas. Potentially, this enables the decoding of additional data messages, whereby the iterative SIC procedure continues.
\end{enumerate}

We compare C-RAPiD with two baseline schemes: scheduled Massive MIMO (SMM) and ALOHA. SMM relies on fully scheduled transmissions and thus 
serves as an upper bound. ALOHA is the classical approach where users randomly select a pilot sequence and a time slot and only collision-free transmissions contribute to the throughput.

\begin{figure}[t]
 \centering
 \includegraphics[width=1\columnwidth]{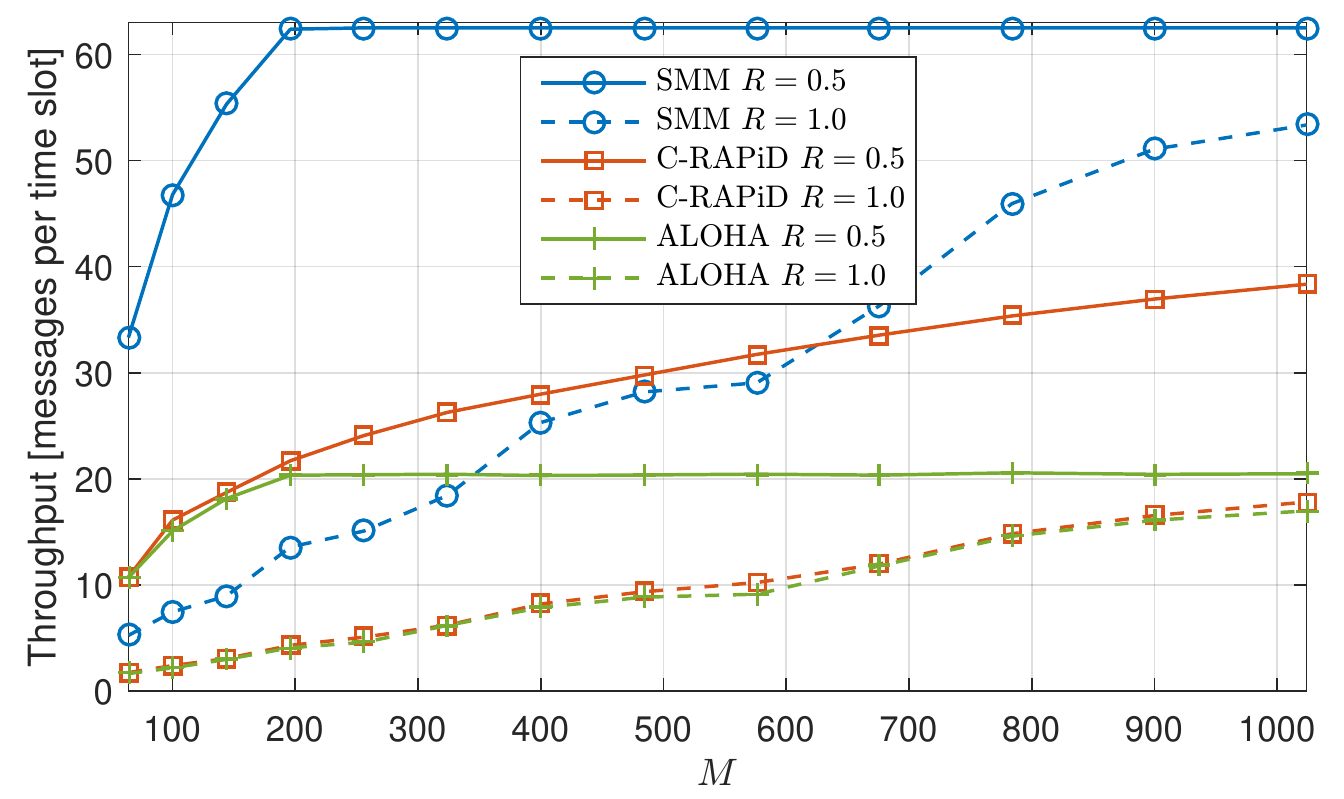}
 \caption{Comparison of throughputs at $R=0.5$ and $R=1$ and optimized
   values of $\tau_p$, $p_a$ and the frame size.}
 \label{fig:gamma_M_schemes}
\end{figure}

We apply the channel model from Section~\ref{sec:ERAPiD} with power control, such that all devices have an SNR of 10~dB. A channel code is applied with rate $R$ and QPSK modulation. For all protocols, $\tau_p$, $p_a$ and $\Delta$ have been
numerically optimized for maximum throughput. As expected, the performance of all protocols increases with $M$, see Figure~\ref{fig:gamma_M_schemes}. However, for $R=0.5$, the performance of ALOHA saturates at roughly $M=200$, whereas C-RAPiD continues to increase. ALOHA can only benefit from the increased SINR until the point that collision-free signals are decoded with high probability. C-RAPiD is able to benefit further due to the SIC mechanism. C-RAPiD achieves $45\%$ of the throughput of SMM with $M=400$ and $R=0.5$, while ALOHA achieves $33\%$. At $M=1024$, the performance of C-RAPiD is increased to $61\%$.

%%%%%%%%%%%%%%%%%%%%%%%%%%%%%%%%%%%%%%%%%%
\section{Conclusions and Perspectives}

Massive MIMO is currently one of the most compelling technologies for 5G
wireless networks.  The operation in TDD and the resulting
uplink-downlink reciprocity renders the system entirely scalable with
respect to the number of BS antennas, leaving channel
coherence (device mobility) as the only remaining, fundamental
limiting factor.  Fast-moving devices result in short coherence and
room for fewer orthogonal pilots in each cell.

In this article, we have addressed crowd-eMBB and mMTC scenarios in which
there exist many more devices in a cell than there are unique
orthogonal pilots, and where devices periodically or sporadically
want to access the network, without prior coordination with the BS.  
Specifically, we saw how the abundance of spatial degrees of
freedom, and the presence of channel hardening, in Massive MIMO
facilitates efficient resolution to resolve colliding transmissions, even in
case the colliding packets use the same pilots. This brought the
central conclusions of the article:
\begin{itemize}
\item Massive MIMO is a fundamental enabler for crowd-eMBB
  scenarios, sensor networks, IoT and M2M communications;
\item The creation of an efficient standard for wireless
  networks based on Massive MIMO technology will require a complete
  re-design of the multiple-access layer.
 \item The spatial domain provides new resources for collision resolution, that are unused in legacy systems.
\end{itemize}

%\bibliographystyle{IEEEtran}
%\bibliography{IEEEabrv,references}


\end{document}